\documentclass[fleqn,twoside,twocolumn,nofootinbib,showkeys]{revtex4} %
\usepackage[sec,nocpr,nopacs]{ujp} 

\newcommand{\beq}{\begin{equation}}
\newcommand{\eeq}[1]{\label{#1} \end{equation}}

\begin{document}
\title[From Euclid to BGL]
{FROM EUCLID TO BGL\,$^1$}%
\author{L\'aszl\'o Jenkovszky}
\affiliation{Bogolyubov Institute for Theoretical Physics, Nat. Acad. of Sci. of Ukraine}
\address{14b, Metrolohichna Str., Kiev 03143, Ukraine}
\email{jenk@bitp.kiev.ua}

\udk{539}  \razd{\seci}

\autorcol{L.\hspace*{0.7mm}Jenkovszky}
\setcounter{page}{977}%

\begin{abstract}
The emergence  of the new, non-Euclidean geometry of Bolyai,
Gau{\ss}, and Lo\-ba\-chev\-skii (BGL) and its impact on modern
sciences is the subject of a series of biennial
conferences.\,\,Be\-low, I briefly review the history.
\end{abstract}

\maketitle

\section{Prologue} \label{s1}
Over  2300 years ago, the great Greek mathematician Eukleides (${
E}\upsilon \kappa \lambda \varepsilon \acute{\iota} \delta \eta
\zeta $ in Greek, Euclides in Latin; in modern literature, the name
is usually written as Euclid, with ``c''; I follow the latter
spelling) from Alexandria (now Egypt) has laid down the foundation
of the geometry now known from textbooks.\,\,It was based on a
number of postulates and axioms.\,\,All but one were generally
accepted either as obvious or logically
consistent.\,\,Ex\-cep\-tional was the 5-th postulate about parallel
lines.\,\,The proof of this postulate has been an embarrassment ever
since Euclidean geometry was founded, although the geometry was not
questioned.\,\,It was accepted, among others, by Isaac Newton,
Leonardo da Vinci, Galileo Galilei, Johann Kepler, Joseph-Louis
Lagrange, and Immanuel Kant.

It was only in the first half of the XIX-th century that three great
men, J.\,\,Bolyai, J.C.F.\,\,Gau{\ss}, and N.L.~Lo\-ba\-chev\-skii
(BGL, alphabetically, Fig.~1) si\-mul\-ta\-neous\-ly and
independently revealed the truth.

These developments and the personal dramas of their protagonists~--
BGL, Fig.~1, motivated the organization of a conference on
Non-Euclidean Geometry and its implications in physics and
mathematics in 1997 in Uzhgorod, Ukraine.\,\,The success of the
conference inspired the organizers to continue: BGL became a series
of biennial conferences, held at various places connected with the
names of the founders of the New Geometry.\,\,The main organizer of
the series of conferences was the Bogolyubov Institute for
Theoretical Physics, National Academy of Sciences of
Ukraine.\,\,Among its founders and active supporters were
Academician Istv\'an Lovas (Budapest), Professors Elem\'er Kiss
(Marosv\'as\'arhely), N.A.~Chernikov (Dubna) and G.M.~Polotovskii
(N.~Novgorod)~--  experts on the subject.\,\,The history of the
series can be traced at:
https://indico.cern.ch/event/586799/page/8964-former-bgl-conferences
and in Ref.\,\,\cite{history2}.\,\,The success of this series of
meetings to a large extent is due to the right choice of the
subject, the cast and style (``key'') of the conferences uniting
physics, mathematics, history, and relevant people coming both from
East and West.\vspace*{-2mm}

\section{Predecessors}
\footnotetext[1]
           {This work is based on the results presented at the
           XI Bolyai--Gauss--Lo\-ba\-chev\-skii (BGL-2019) Conference: Non--Eucli\-de\-an,
           Noncommutative Geometry and Quantum Physics.}By mentioning the fifth postulate last, Euclid himself, in this way,
alluded to a deficiency.\,\,A great number of Greek, Arabic,
Renaissance, and other mathematicians tried to prove, disprove,
generalize, or replace the postulate under question.\,\,In\-te\-rest
in geometry was enhanced in the 17-th and 18-th centuries preparing
the great harvest in the 19-th century.

Before BGL, at least two men came close to the concept of the new
geometry.\,\,One was Girolamo Saccheri, an Italian monk, who made
the right step in his ``Logica demonstratica'' to resolve the
contradiction.\,\,The idea was further elaborated in his paper
entitled ``Euclides ab omni naevo vindicates'' (``Euclides free of
any shadow''), published in 1697.\,\,Saccheri's work did not remain
unnoticed: it became familiar, e.g., in G\"ottingen due to the
thesis of Kl\"ugel (a student of Prof.\,\,A.~K\"ostner) ``Conatuum
praecipurum theoriam parallelarum demonstrandi recensio''), where it
was thoroughly reviewed.\,\,Later on, in 1766, Saccheri's approach
was
further developed by \mbox{Lambert.}

Between  1807 and 1816, Schweikert, a German lawyer in Kharkov ({\it
sic!}), developed his version of non-Euclidean geometry, called
``Astralische Geometrie'' (alluding to cosmic scales at which any
departure from the Euclidean geometry may be
noticeable).\,\,Schwei\-kart, an amateur mathematician, did not use
any formalism, his ideas were mathematically formalized by his
nephew Taurinos, who in 1826 published his ``Geometria prima
elementa'', in which the ``log-spherical'' formalism, preceding that
of Bolyai and Lo\-ba\-chev\-skii was used to prove Euclid's 5-th
\mbox{postulate.}

Gau{\ss} was familiar with the work of Schweikert and Taurinos.

\section{Bolyai, Gau{\ss}, and Lo\-ba\-chev\-skii (BGL)}

 \begin{figure}
\vskip1mm
\includegraphics[width=\column]{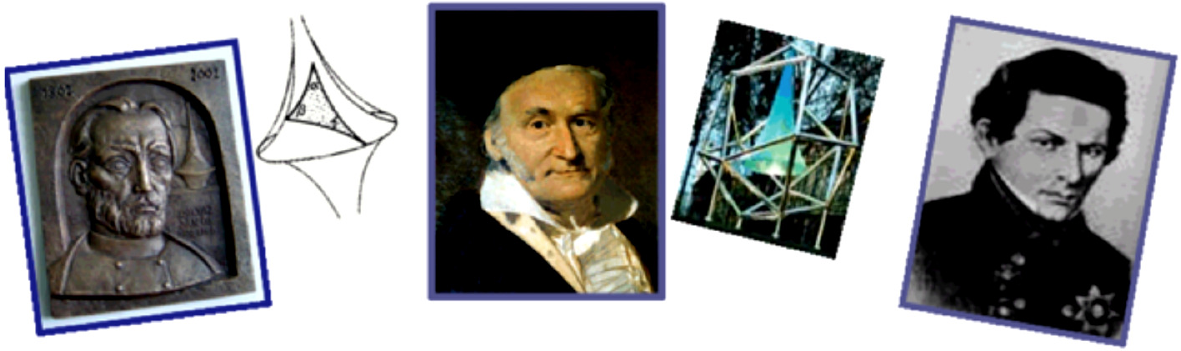}
\vskip-3mm\caption{J. Bolyai, J.C.F. Gau{\ss}, and N.I.
Lo\-ba\-chev\-skii}
    \label{Fig:1}
\end{figure}

 \begin{figure}
 \vskip1mm
\includegraphics[width=\column]{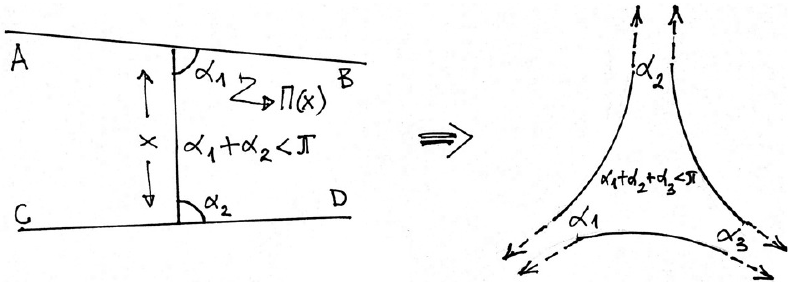}
\vskip-3mm\caption{Visualizing the non-Euclidean geometry}
    \label{Fig:2}\vspace*{-2mm}
\end{figure}

In  the new geometry, the sum of internal angles is not $\pi$ any
more (Fig.~2).\,\,It can be smaller depending on the length of the
sides.\,\,The new ``parallelism angle'' $\Pi(x)$ (equivalent of
$\pi/2$ in the Euclidean case) is related to the distance $x$, see
Fig.~2, by
\begin{equation}
{\rm ctg}\Pi(x)/2=q^x,
\end{equation}
where $q$ is a parameter.

The  curvature in the right panel of Fig.~2 is accentuated
(enhanced) for pedagogical reasons.\,\,J.~Bo\-lyai, C.F.~Gau{\ss},
and N.I.~Lo\-ba\-chev\-skii were aware of the non-observability of
any departure from the Euclidean geometry within the visible
Universe.\,\,This yields the epithet ``new'' (``absolute'',
``imaginary'', ``pan'',\,\,...)\,\,and the reason why contemporaries
were so re\-luctant to accept the apparently abstract
\mbox{construction.}

It is like a mystery how a problem open for millennia could have
been resolved practically simultaneously (within a decade)
independently by three men who never met and did not communicate,
see their (simplified) world lines, last Section.

\subsection{Johann Carl Friedrich Gau{\ss} }
The  eldest among BGL was born in April 30, 1777 in Braunschweig
\cite{Buhler}.\,\,In 1795, he entered the G\"ottingen University,
where Farkas Bolyai (J\'anos' farther) became his closest friend
during his studies.\,\,After 3 years in G\"ottingen, Farkas returned
to his homeland Transylvania, where he became a teacher of
mathematics in Marosv\'as\'arhely.\,\,The friendship between Farkas
and Carl Friedrich however lasted for decades with an extensive
exchange of letters, providing valuable information for the history
of science.

In 1804, Farkas sent his ``proof'' of the 5-th postulate to
Gau{\ss}.\,\,In his reply, Gau{\ss} indicated an error in the
derivation of his friend, adding that himself he also hopes to
progress in solving the problem.\,\,It looks surprising that
Gau{\ss}, commenting Lo\-ba\-chev\-skii's ``Geo\-met\-ri\-sche
Un\-ter\-su\-chun\-gen'', wrote to Schu\-ma\-cher in 1846 that he
found in Lo\-ba\-chev\-skii's work ``nothing new''.

In 1815 in a comment on M\"atternik's book, he wrote: ``...we should
admit that we are unable to advance compared to the 2 thousands
years old Euc\-lides''.\,\,Mo\-reo\-ver, in a letter to Olbers
(28.04.1817), he wrote that ``geometry cannot be proven by human
intelligence''.

Schweikert published his paper on parallels ``Astralische
Geometrie'' in 1807 and developed it further after he moved from
Kharkov to Marburg.\,\,Gau{\ss}' work was inspired to a large extent
by a letter of Schweikert in 1819, in which Schweikert informed
Gau{\ss} of his geometry.\,\,Tau\-ri\-nos continued his work in 1924
in close touch with Gau{\ss}.\,\,As mentioned above, Schweikert and
Taurinos stopped half-way, unable to abandon the Euclidean way of
reasoning.

After 1816, Gau{\ss}, partly inspired by practical goals, started
working on geodesics, and, in 1828, he published his famous paper on
differential geometry (curved spaces).\,\,Ho\-we\-ver, it was only
E.\,\,Belt\-rami who in 1868  suggested the interpretation of
non-Euclidean geometry in terms of surfaces with negative curvature.

In 1832, Gau{\ss} became aware of the work by J\'anos Bolyai,
mediated by Farkas, and he acknowledged Lo\-ba\-chev\-skii's
``Geometrische Untersuchnungen'' in 1840.\,\,The same year, he
started learning Russian; A.S.~Pushkin's ``Boris Godunov'' in
original Russian was found in his library.

Gau{\ss}  did not publish a single paper on non-Euc\-li\-dean
geometry.\,\,On various occasions, e.g., in his letters, he
privately praised both Lo\-ba\-chev\-skii and J\'a\-nos Bo\-lyai for
their contributions in developing the new geometry, but he never
cited them publicly for that achievement! In 1842,
Lo\-ba\-chev\-skii was nominated Member of the G\"ottingen
Scientific Society. Ho\-we\-ver, in Gau{\ss}' recommendation,
Lo\-ba\-chev\-skii's work on geometry was not mentioned.

\subsection{Nikolai Ivanovich Lo\-ba\-chev\-skii }

was  born in Nizhni Novgorod (where N.N.\,\,Bo\-go\-lyu\-bov was
also born, by the way) on November 20, 1792 \cite{Laptev}.\,\,His
studies and professional carrier are connected with  Kazan,  next
big town downstream \mbox{Volga.}\looseness=1

The first presentation of his ``new geometry'' took place at the
Department of physical and mathematical sciences of the Kazan
University on February 7, 1826.\,\,A formal application of the
Department asking to publish his presentation entitled ``Exposition
succinte des princeples de la geometrie'' (in French) was rejected
by the local ``Uchenye zapiski''.\,\,The original manuscript was
lost.

In 1827, N.I.\,\,Lobashevskii was elected Rector  of the Kazan
University.

The first  publication of the new geometry is dated 1829, when
``Kazanskii Vestnik'' published Lo\-ba\-che\-vskii's ``On the
principles of geometry'' (in Russian).\,\,In 1932, this paper was
submitted to the Russian Academy of Sciences (in
St.\,\,Pe\-ters\-burg).\,\,It was reviewed and rejected by
M.V.~Ostrogradskii, whose report was totally negative.\,\,Moreover,
the magazine ``Syn otechestva'' published an ironic anonymous
pamphlet in 1834 brutally criticizing the author of the new
geometry.

These misfortunes did not discourage Lo\-ba\-chev\-skii.\,\,He
continued writing and publishing, ``Geomet\-ri\-shce
Untersuchungen'', in German, among others.\,\,Ul\-ti\-ma\-tely, one
year before his death, ill and blind, he dictated his
``Pangeometry'', published in 1855 in Russsian, followed by the
translation into French and publication in 1856.\,\,The Bolyais and
Gau{\ss} became familiar with the ``Geometrishce
Untersuchungen''.\,\,Far\-kas, in his book of 1851 and Gau{\ss} (in
a private letter) appreciated and praised it, while the author was
still alive.\,\,Nevertheless, Lo\-ba\-chev\-skii never received
public recognition during his life.\,\,He died ill, on February 12,
1856, in misery.

\subsection{J\'anos Bolyai\,$^2$}
was  the youngest and maybe the most tragic personage among BGL.
Born in Kolozsv\'ar  (1802), he moved in 1804 with his father Farkas
to Marosv\'as\'arhely, both in Transylvania\,\footnotetext[2]{In
Hungarian, contrary to other European languages, the family name
precedes the given name: Bolyai J\'anos.}  (now Rumania), see the
map in Fig.~3.

\begin{figure}
\vskip1mm
\includegraphics[width=7.5cm]{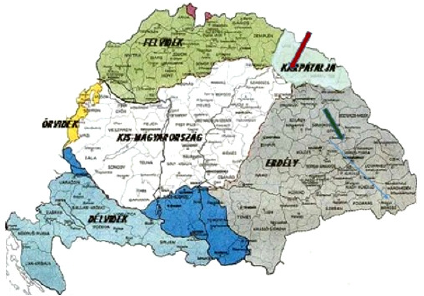}
\vskip-3mm\caption{Hungary of Bolyai's time. The green
line points to Bolyai's homeland, and the red
one points to Uzhgorod (Ungv\'ar), site of the first BGL
meeting in 1997. Today's Hungary is in white}
    \label{Fig:3}
\end{figure}

In 1818, Farkas asked his friend C.\,\,Gau{\ss} to support further
studies of J\'anos at the G\"ottingen University.\,\,But, for some
reason, Gau{\ss} was reluctant to do so.\,\,Consequently, J\'anos
went to the Engineering Academy in Vienna.\,\,After  graduation, he
was appointed a custom officer in Temesv\'ar.

In 1820, he informed his father that he found the way to prove
Euclid's 5-th postulate.\,\,Farkas discouraged his son from doing
something he considered hopeless.

Despite father's advise, J\'anos continued his efforts, and, in
1824, he found a mathematical relation between the length of a
perpendicular and the angle of the asymptote.\,\,``I created a world
from nothing'', he exclaimed to his father.\,\,One can see from his
manuscripts, sketches, and letters that, already in 1820, he was on
the right track in considering the limit of the large
circle.\,\,Most of the work done by J\'anos Bolyai remained
unpublished.\,\,Ma\-nu\-scripts can be found \cite{Kiss} in
libraries and museums, e.g., in Marosv\'as\'arhely.

In February 1825, J\'anos sent his manuscript to his
farther.\,\,Farkas was unable to abandon the old geometry, whose
proof took part of his life, in vain.\,\,Re\-luc\-tant to follow the
ideas of his son, he was looking for possible shortcomings in the
work of J\'anos.\,\,Finally, in February 1829, he agreed to publish
the results of his son as an appendix to  his own book ``Tentamen
Juventutem...'', a mathematical course for young people.\,\,The
appendix (in Latin) was entitled ``Appendix scientium spatii
absloute veram exhibeus''.\,\,The book appeared in 1831.\,\,A copy,
immediately sent by Farkas to Gau{\ss} did not reach the recipient:
the area was plagued by cholera.\,\,Another copy had reached the
designation at the beginning of 1832.\,\,Gau{\ss} reacted
immediately, in March 1832.\,\,The response was fatal for J\'anos.
``You may be surprised~-- he wrote~-- that I will not praise your
son's work since praising it would mean praising myself...\,\,his
ideas almost coincide with my way of thinking during 30--35 years
from now...\,\,Myself, I also intended to publish these results, but   
once my friend's son did it, I feel free from that duty''.\,\,The
above letter was preceded by another one, to Gerling in which
Gau{\ss} praises J\'anos as a first-rank genius.

Gau{\ss}'  letter to Farkas made happy the farther, but not his
son.\,\,J\'anos suspected that his ideas were
stolen.\,\,Fur\-ther\-more, in 1848, he received
Lo\-ba\-chev\-skii's ``Geometrische Untersuchungen''.\,\,He even
suspected that ``Lo\-ba\-chev\-skii'' is a pseudonym used by
Gau{\ss} as a mask.\,\,Af\-ter the first shock and the resulting
depression, he started critically reading Lo\-ba\-chev\-skii's work
that was really very close in spirit to what J\'anos did.

Bolyais' native language, used in communication and correspondence,
was Hungarian, although their scientific works were written in Latin
or German.\,\,J\'a\-nos was fluent also in Italian and French and
was familiar with Chinese and Tibetian.\,\,He was working  on the
idea to reform the Hungarian language, aiming to adapt it to
scientific texts.\,\,(NB: The precondition to a candidate to the
Hungarian Academy of Sciences was a publication in Hungarian).\,\,He
believed that Hungarian, due to its particular grammar, is a perfect
basis for a future universal scientific language
\cite{Marac}.\,\,Its basis should be semantic
(symbolic).\,\,J\'a\-nos was obsessed with the so-called ``root
words'' (gy\"okszavak) or homonimas, unique for the Hungarian.\,\,He
argued so by comparing Hungarian with Latin and German, e.g., in the
phrase:

           P\'eter ember

           Petrus est homo

           Peter ist ein Mensch.

He was trying to fix, as in mathematics, ambiguities, i.e., to
establish a one-to-one correspondence between words
(symbols~=~semiotics) and notions. There is, however, a controversy
between  simple words-symbols and the complicated Hungarian grammar
(e.g., in conjugations and declinations).\,\,(See
Ref.\,\,\cite{Marac} for further reading on this subject.)\,\,A
relatively new development in this direction is connected with the
use of modern communication means (computers, Internet, e-mail,
\textit{etc.}), where new (telegraphy) languages are being developed
automatically (shorthand, neglect of accents, ``likes'',
\textit{etc.}).

J\'anos was also an excellent violin player and good fencer.

His legacy consists of over 15 000 pages of ma\-nu\-scripts, written
in special codes, stored in the Te\-le\-ky library in
Marosv\'as\'arhely.\,\,Elem\'er Kiss con\-tri\-bu\-ted \cite{Kiss}
largely to their de-co\-di\-fi\-ca\-tion.\,\,It is not always clear
whether the texts imply shorthand-wri\-ting  (stenography) or
homonimas.\vspace*{-1mm}

\section{Followers}
The  next breakthrough came with the syntheses of non-Euclidean and
differential geometry with surface theory, resolving the problem of
uniqueness of the new geometry~-- the main obstacle on the way to
its approval.\,\,Met\-rics, geodesics, curvature \textit{etc.}
provided classification and interpretation of the new geometry.

Berngard Riemann  (1826--1866), with his 1854 inauguration lectures
at the G\"ottingen University, published in 1854, introduces the
notion of manifolds replacing the space, with points corresponding
to elements of a manifold.\,\,The geometry of the manifold is
defined by the squared distance between infinitesimally close
points:\vspace*{-1mm}
\begin{equation}
ds^2=\sum_{i,j=1}^nq_{ij}dx_idx_j,
\end{equation}
where $q_{ij}(x)$ is the metric tensor.

The  above equation defines the Riemann metrics within the geometry
of the Riemann space.

Geometries  are classified according to their curvatures $K$: $K>0$
is called Riemann (or elliptic) geometry, $K=0$ is for the
conventional Euclidean geometry, and $K<0$ corresponds to the
non-Euclidean, BGL geometry.

Lorentz,  Poincar\'e, Minkowski and Einstein contributed to making
the new geometry a common physical language.\,\,The four-dimensional
Lorentz space-time and the pseudo-Euclidean metrics
\begin{equation}
ds^2=dx^2+dy^2+dz^2-c^2dt^2,\quad K=0
\end{equation}
form  the basis of modern physics, in particular of the  relativity
theory, resulting, e.g., in the relation between mass and energy
\begin{equation}
E=mc^2,
\end{equation}
with global consequences for the mankind.

Cosmology and the history of our Universe are al\-so based on the
new geometry.\,\,So\-viet physicist A.A.~Fried\-man (1888--1925) has
found a special solution of the Einstein equation within the
Bo\-lyai--Lo\-ba\-chev\-skii metrics in 1922 and predicted the
expansion (in\-fla\-tion) of our Universe confirmed in 1929
\mbox{(Hubble)}.

The  new geometry has enriched natural science conceptually,
although the curvature effects are negligible within observable
distances.\,\,Lo\-ba\-chev\-skii suggested tests based on the
measured angular distances (parallax) of stars to the astronomer
Struve.\,\,He proposed also the calculation of definite integrals on
surfaces and volumes extended to infinity.\,\,About $200$ integrals
calculated in this way are available in textbooks and tables.

Interesting are applications in architecture using hyperbolic constructions, see
\cite{Hyper}.

A  modern branch of the non-Euclidean geometry is connected with the
so-called quantum groups or $q$-deformations.

\section{Epilogue}
The abrupt birth, at the same time but different places, of the new
geometry, after thousands dormant years, seems almost a
mystery.\,\,My\-ste\-rious is also the fate of its creators,
thinking almost identically, while living at the same period on the
same continent, without knowing about each other (see Fig.~4).

\begin{figure}
\vskip1mm
\includegraphics[width=\column]{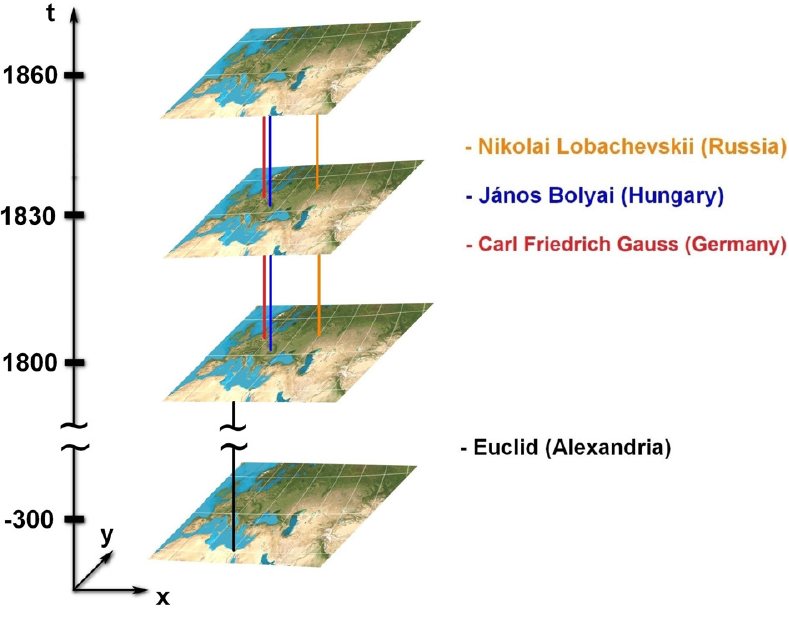}
\vskip-3mm\caption{Time geography: simplified (straightened) world
lines of Euclid, J\'anos Bolyai, Gau{\ss}, and Lo\-ba\-chev\-skii.
Their parallels never crossed}
    \label{Fig:4}
\end{figure}

The BGL  conferences usually include review talks and discussions on
history and biographies of Bolyai, Gau{\ss}, and
Lo\-ba\-chev\-skii, their relation to teachers and followers.\,\,The
priority of the discovery was never questioned at the
conferences.\,\,All three deserve appreciation.\,\,Both J\'anos
Bolyai and N.I.~Lo\-ba\-chev\-skii, for various reasons, were
unfortunate in failing to produce a timely open publication on their
discoveries.\,\,Ho\-we\-ver, this fact cannot justify any dispute on
priorities.\,\,Gau{\ss} did not publish a single paper on
non-Euclidean geometry, still his reputation is so high that nobody
doubts that he ``knew it''.\,\,A rather unique situation in the
history of science?! The alphabetic ordering of BGL at conferences
was accepted from the very beginning of the series and was never
questioned, which, however, did not prevent the speakers to have
personal preferences and diverse points of~view.

The previous BGL conferences were dominated by Hungarian and
Ukrainian-Russian-Bielorussian organizers and participants, extended
by  wide international participation.\,\,Co-patriots of Gau{\ss} are
welcome at future BGL meetings! The BGL series has a certain Middle
European (Mittleurop\"aische) flavor enriched by participants from
far-away countries. The next BGL conference will be held in Lviv
(Western Ukraine) on July 4--9, 2021, visit:
http://indico.bitp.kiev.ua/e/bgl-2021.

\vskip3mm \textit{I had pleasure and profited very much from
communication with participants of BGL meetings, where I knew famous
personalities like A.N.~Bogolyubov, outstanding expert in history of
science, brother of Nikolai Nikolaevich, Istv\'an Lovas, Elem\'er
Kiss, and N.A.~Chernikov and discovered new worlds such as
Transylvania and the Volga area~-- hosts of BGL conferences and
fertile birthplaces of great individuals.}

\textit{I thank the organizers of the present meeting for inviting
me to present this talk.}


\vspace*{-8mm}
\rezume{%
Л.Л.\,Єнковський}{ВІД ЕВКЛІДА ДО БГЛ} {Поява нової неевклідової
геометрії Бояі, Гауса і Лобачевського (БГЛ) та її вплив на сучасну
науку стала предметом вивчення серії дворічних конференцій. В цій
роботі я коротко нагадую її історію.}


\begin{thebibliography}{99}
\bibitem{history1}
https://indico.cern.ch/event/586799/page/8964-former-bgl-conferences\,.

 \bibitem{history2} L.L. Jenkovszky. {\it BGL conferences: a brief history,
Acta Physica Debrecina, TOMUS XLII, Redigid Zolt\'an Tr\'ocs\'anyi,
2008}.

\bibitem{Laptev} \textit{Nikolai Ivanovich Lo\-ba\-chev\-skii} (Kazan' University, 1979) (in Russian).

\bibitem{Buhler} W.K. B\"{u}hler. {\it Gau{\ss}. A Biographic Study} (Springer,  1981).

\bibitem{Kiss} Kiss Elem\'er. {\it Matematikai Kincsek Bolyai J\'anos K\'eziratai Hagyat\'ekaibol} (Akad\'emiai kiad\'o, 1999) (available also in English).

\bibitem{Marac} Mar\'acz S\'andor, {\it Bolyai J\'anos \'es a Magyar Mint T\"ok\'eletes Nyelv}. http://kincseslada,hu/magyarsag/con\-tent.php?article.328\,.

\bibitem{Hyper} https://indico.cern.ch/event/586799/contributions/2695991/ attachments/1512907/2359928/Jenkovszky.pdf\,.

\begin{flushright}
{\footnotesize Received 10.09.19}
\end{flushright}
\end{thebibliography}
\end{document}